\documentclass[12pt]{article}
\usepackage{amssymb}
\usepackage{amsbsy}
\usepackage{epsfig}
\usepackage{verbatim}
\makeatletter \@addtoreset{figure}{section}
\def\thefigure{\thesection.\@arabic\c@figure}
\def\fps@figure{h, t}
\@addtoreset{table}{bsection}
\def\thetable{\thesection.\@arabic\c@table}
\def\fps@table{h, t}
\@addtoreset{equation}{section}

\newtheorem{corollary}{Corollary}[section]
\newtheorem{definition}{Definition}[section]
\newtheorem{theorem}{Theorem}[section]
\newtheorem{proposition}{Proposition}[section]

\newtheorem{lemma}{Lemma}[section]
\newtheorem{remark}{Remark}[section]
\newtheorem{remarks}[remark]{Remarks}

\def\bd{\begin{definition}}
\def\ed{\end{definition}}
\def\bt{\begin{theorem}}
\def\et{\end{theorem}}
\def\bp{\begin{proposition}\rm}
\def\ep{\end{proposition}}
\def\bc{\begin{corollary}}
\def\ec{\end{corollary}}
\def\bl{\begin{lemma}\em}
\def\el{\end{lemma}}
\def\be{\begin{equation}}
\def\ee{\end{equation}}
\def\br{\begin{remark}\rm\small}
\def\er{\end{remark}}
\def\brs{\begin{remarks}.\\ \rm\
\begin{enumerate}}
\def\ers{\end{enumerate}\end{remarks}}
\def\bea{\begin{eqnarray}}
\def\eea{\end{eqnarray}}

\def\Tr{\mathrm {Tr}}
\def\tr{\mathrm {tr}}
\def\det{\mathrm {det}}

\def\diag{\mathrm {diag}}

\def\&{&{\hskip -20pt}}

\def\pb{\mathbf{p}}

\def\nchi{\hbox{\raise 2.5pt\hbox{$\chi$}}}
\date{}
\begin{document}
\baselineskip 16pt
\medskip
\begin{center}
\begin{Large}\fontfamily{cmss}
\fontsize{17pt}{27pt}
\selectfont
\textbf{SU(N) integrals and tau functions}
\footnote{}
\end{Large}\\
\bigskip
\begin{large}  
 {A. Yu. Orlov}$^{1,2}$
 \end{large}
\\
\bigskip
\begin{small}
$^{1}${\em Nonlinear Wave Processes Laboratory, \\
Oceanology Institute, 36 Nakhimovskii Prospect,
Moscow 117851, Russia\\
 e-mail: orlovs@ocean.ru } \\ 
$^{2}${\em National Research University Higher School of Economics, \\, Moscow, Russia}\\
\end{small}
\end{center}
\bigskip

\begin{center}{\bf Abstract}
We present a family of solvable multi-matrix models associated with an arbitrary embedded graph  
$\Gamma$
with a single vertex. The graph with $n$ edges is equipped with $2n$ corner matrices. The partition function of each member of the family depends on the set of eigenvalues of monodromies of corner matrices around the vertices of the dual graph 
$\Gamma^*$ and set of parameters attached to the vertex of $\Gamma$. We select the cases where the partition function of a model is a tau function of KP, 2KP and BKP hiearachies. The answer can be written as a determinant in the KP case and a pffafian in the BKP case of a matrix whose entrence consists of generalized hypergeometric functions. We compare integrals over ${U}(N)$ and over
${SU}(N)$ groups. In $U(N)$ case there is no restriction on the number of vertices of $\Gamma$.
We also consider mixed ensembles of matrices from $GL(N),U(N)$ and $SU(N)$.
\end{center}
\smallskip

\begin{small}

 \bigskip
\end{small}
\bigskip \bigskip

\section{Introduction}

There is a  number of detailed studies of one- and two-matrix models on $SU(N)$ and mainly 
on $U(N)$ cases. I think it is difficult to give a complete set of reference, let us mention 
\cite{HarishChandra},\cite{ItzyksonZuber},\cite{Shatashvili}, \cite{MirMorSemen}, \cite{Mehta},\cite{ZinnZuber},\cite{TracyWidom}, \cite{Zuber2016}, \cite{KMMOZ} \cite{BrezinGross},\cite{GrossWitten}. In particular, $SU(N)$ case was studied in \cite{Creutz},\cite{Collins},\cite{Carlsson}.

As for multi-matrix models, the pioneer work was \cite{Witten91}; then see \cite{Kostov},\cite{Aleksandrov},\cite{NO2020},\cite{NO2020tmp}, \cite{AOV}. 
In what follows we use the construction \cite{AOV}.

\paragraph{Preliminaries.} Here, we gather facts we need in what follows, using
mainly \cite{AOV} and \cite{BChVol}.

We can write the Haar measure $d\mu(U)$ of the ${SU}(N)$ group as the product of the Haar measure 
$d_*U$ over ${U}(N)$ group times Dirac delta function $\delta\left(\arg[\det U-1]\right)$, where
\be\label{SUcorrection}
\delta\left(\arg[\det U- 1]\right)=(2\pi)^{-N}\sum_{q\in Z} e^{q(\theta_1+\cdots + \theta_N)}
=(2\pi)^{-N}\sum_{q\in Z} \det\, U^q
\ee
Thus, one can write
\be\label{int=int}
\int_{{SU}(N)} f(U)d\mu(U)=\int_{{U}(N)}f(U)\delta\left(\arg[\det U -1]\right)d_*U
\ee

Having this relation in mind one can conclude that the relation (see Appendix \ref{derivation})
\be\label{2'}
\int_{{U}(N)}
s_\lambda(UAU^\dag B) \det\,U^q \,d_*U = \frac{s_\lambda(A)s_\lambda(B)}{s_\lambda(I_N)}\delta_{q,0}
\ee
(where $s_\lambda(X)$ is the Schur function labeled by a patition $\lambda$ and where $A$ and $B$ are
arbitrary complex matrices) is also valid for ${SU}(N)$ if we replace $d_*U$ by $d\mu(U)$:
\be\label{3'}
\int_{{SU}(N)}
s_\lambda(UAU^\dag B) d\mu(U) =
\int_{{U}(N)}
s_\lambda(UAU^\dag B) d_*U = \frac{s_\lambda(A)s_\lambda(B)}{s_\lambda(I_N)}
\ee

At the same time from
\be\label{2a}
\int_{{U}(N)} s_\lambda(UA)s_\mu(U^\dag B)\,d_*U=
\delta_{\mu,\lambda}\frac{s_\lambda(AB)}{s_\lambda(I_N)}
\ee
and from 
$$
s_\lambda(UA)\,\det\, U^q=s_{\lambda+q^N}(UA)\,\det\,A^{-q},\quad q>0
$$
where $\lambda+q^N$ denotes the partition  $(\lambda_1+q,\lambda_2+q,\dots)$
and from 
$$
s_\mu(U^\dag B)\,\det\, U^q=s_{\mu+|q|^N}(U^\dag B)\,\det\,B^{-|q|},\quad q<0
$$
we have
\be\label{2b}
\int_{{U}(N)} s_\lambda(UA)s_\mu(U^\dag B)\,\det U^q \,d_*U=
\delta_{\mu,\lambda+q^N}\frac{s_\mu(AB)}{s_\mu(I_N)}\det A^{-q},\quad q>0
\ee
\be\label{2c}
\int_{{U}(N)} s_\lambda(UA)s_\mu(U^\dag B)\,\det U^q \,d_*U=
\delta_{\lambda,\mu+q^N}\frac{s_\lambda(AB)}{s_\lambda(I_N)}\det B^{q},\quad q< 0
\ee

It results that for ${SU}(N)$ case we get:
\[
\int_{{SU}(N)} s_\lambda(UA)s_\mu(U^\dag B)d\mu(U) =
\delta_{\mu,\lambda}\frac{s_\lambda(AB)}{s_\lambda(I_N)} +
\]
\be\label{3bc}
+ \sum_{q>0} \delta_{\mu,\lambda+q^N}\frac{s_\mu(AB)}{s_\mu(I_N)}\det A^{-q}+
\sum_{q<0}\delta_{\lambda,\mu+q^N}\frac{s_\lambda(AB)}{s_\lambda(I_N)}\det B^{q}
\ee
which is actually written down in \cite{BChVol} in a little bit different form.

Thus, we get
\be\label{4}
\int_{{SU}(N)} s_\lambda(UA)s_\lambda(U^\dag B)d\mu(U) =
\frac{s_\lambda(AB)}{s_\lambda(I_N)} 
\ee

Now, we shall see that the pair of relations (\ref{3'}) and (\ref{4}) allows
to constract multimatrix models associated with any embedded graph with a single vertex.
It also allows the equate partition function of composed matrix models based on mixed ensembles
which consists of complex, unitary and special unitary matrices, to tau functions.

\section{Multi-matrix models}

We begin with the construction of multi-matrix models according to \cite{AOV},
 \cite{NO2020tmp}, \cite{AOV}). Here, we present it in a little bit different way.
Subsection \ref{new} contains new results.

\subsection{Graphs, corner matrices, integrals of the Schur functions}

\paragraph{Graphs. Dual,
graphs. Corner matrices.}
An embedded graph is a graph drawn on a closed oriented surface without boundary, where
each face of the graph is homeomorphic to a disk.
In this case, the edges are ribbon edges.
For an embedded graph, say $\Gamma$, the sum of the number of vertices $V$ minus
the sum of the number of edges $n$ plus the sum of the number of faces $F$ is equal to the Euler characteristic of the surface on which the graph is drawn.

Recall that the graph $\Gamma^*$, dual to $\Gamma$, has $V$ faces, $n$ edges, and $F$ vertices. Every edge
of $\Gamma^*$ intersects an edge of $\Gamma$. Every face of $\Gamma^*$ contains one vertex
of $\Gamma$, and vice versa: every face of $\Gamma$ contains one vertex
of $\Gamma^*$.

Let us number the edges with positive integers $1,\dots,n$, and assign the sides
of edge $i$ the numbers $i$ and $-i$, where it doesn't matter which side is numbered $i$ and which $-i$ (the choice should be recorded).

Let's number the corners around each vertex as follows. If we walk around a vertex
in the positive direction and intersect an edge with number $|i|$, we'll first encounter side $-i$,
and after the intersection, we'll leave side $i$ (here $i$ is taken from the set $\pm 1,\dots, \pm n$).
The corner we've found will be numbered $i$. Let's assign an $N\times N$ matrix $C_i$ to this corner. We call these matrices corner matrices. We have $2n$ corner matrices for a graph with $n$ ribbon edges whose numbers are taken from the set $\pm 1,\dots,\pm n$.

Let's number the vertices.
Now, let's select a vertex $a$ ($a=1,\dots,V$) and multiply corner matrices around the vertex,  according to the positive direction of traversal around the vertex. We'll denote this product
by $\frak{V}_a$ and call it the monodromy of the vertex $a$; monodromy is defined up to a cyclic permutation.

Thus, we have a graph and the set of corner matrices.

Now consider the graph $\Gamma^*$, the dual of $\Gamma$.
Each corner of $\Gamma^*$ faces one of the corners of $\Gamma$, and we number
each corner of the dual graph with the same number. We introduce the monodromies $\frak{V}^*_b$,
$b=1,\dots,F$
of the vertices of $\Gamma^*$ in the same way, but now we traverse the vertices
in the negative direction. Thus, the set of corner matrices $\{ C_{\pm i},\,i=1,\dots,n\}$
produces two sets of monodromies:
$$
\Gamma: \qquad (\frak{V}_1,\dots,\frak{V}_V)\,\longleftrightarrow\, (\frak{V}^*_1,\dots,
\frak{V}^*_F)
$$
We note that the relationship between these two sets is defined by the embedded graph.

We introduce the set of matrices $\{X_{\pm i}\in G,\,i=1,\dots,n\}$ conditioned by $X_{-i}=X_i^\dag$.
In what follows, we consider three cases: $G={U}(N)$, $G={SU}(N)$ and $G={GL}(N)$ and $d\mu(X)$ denotes the related Haar measure for $G={U}(N)$, $G={SU}(N)$
and the Gaussian measure for $G={GL}(N)$ which is defined as
\be\label{Gauss}
d\mu(X^{(c)})=K\prod_{i,j\le N} e^{-N|X^{(c)}_{ij}|^2}d^2 X^{(c)}_{ij},\quad c=1,\dots,n
\ee
where $K$ is the normalization constant chosen by $\int_{{GL}(N)}d\mu(X)=1$, and
$d\mu(X)=\prod_{c=1}^n d\mu(X^{(c)})$. (Here, localy, we wrote $X^{(c)}$ instead of $X_{|c|}$ to
make formula (\ref{Gauss}) more readable).

\paragraph{Dressing. Integrals over $G^{\times n}$, $G=GL(N)$ and $G=U(N)$ cases}

Next we introduce the {\it dressing} procedure:
\be
C_i\,\to \, X_iC_i,\quad i=\pm 1,\dots, \pm n
\ee
which induces the dressing of the monodromies:
\be
\frak{V}_a \,\to\,\frak{V}_a(X),\, a=1,\dots,V;\qquad 
\frak{V}^*_a \,\to\,\frak{V}^*_b(X),\, b=1,\dots,F
\ee

In the case $G={U}(N)$, as well as in the case $G={GL}(N)$ we have
\bp\label{Propos} For $G=U(N)$ and $G=GL(N)$ and
for a set of partitions $\lambda=\lambda^{(1)},\lambda^{(2)},\dots,\lambda^{(V)}$ we get
\be\label{Schurs}
\int_{G^{\times n}} \prod_{a=1}^V s_{\lambda^{(a)}}\left(\frak{V}_a(X)  \right)
d\mu(X)=d_{\lambda,G}^{-n} 
\prod_{a=1}^V\delta_{\lambda,\lambda^{(a)}}
\prod_{b=1}^F s_\lambda(\frak{V}^*_b),
\ee
where
\be
d_{\lambda,{U}(N)}=s_\lambda(I_N),\ee
\be
d_{\lambda,{GL}(N)}=s_\lambda(N\pb_\infty),\quad
N\pb_\infty=(N,0,0,\dots)
\ee
We also have the dual relation:
\be\label{Schurs^*}
\int_{G^{\times n}} \prod_{b=1}^F s_{\lambda^{(b)}}\left(\frak{V}^*_b(X)  \right)
d\mu(X)=
d_{\lambda,G}^{-n} 
\prod_{b=1}^F\delta_{\lambda,\lambda^{(b)}}
\prod_{a=1}^V s_\lambda(\frak{V}_a)
\ee

\ep

 Question: Can relations similar to (\ref{Schurs}) and (\ref{Schurs^*}) be fulfilled for other groups?

For the special unitarty group we have

\bp\label{V=1}
In case $V=1$ we have
\be\label{Schurs_for_SU}
\int_{SU(N)^{\times n}}  s_\lambda\left( \frak{V}_(X)  \right)
d\mu(X)=\left(s_\lambda(I_N)\right)^{-n} 
\prod_{b=1}^F s_\lambda(\frak{V}^*_b),
\ee 
 \ep
 
Without details,
 the proof follows from the complete similarity of (\ref{3'})
with the ``cutting'' relation $(1,-1)\otimes (1,-1)=(1)(-1)$ in the permutation group $S_2$,
acting on the set of two numbers $1$ and $-1$, and the similarity of the  relation (\ref{2a}) with the ``joining'' relation $(1,-1)\otimes (1),(-1)=(1,-1)$ in the same group. Here $(1)$ and $(-1)$ denote cycles of length one, and $(1)(-1)$ is an element of $S_2$ with such a cyclic structure.
Furthermore, $(1,-1)\in S_2$ denotes a cycle of length two. For every element of the group $S_{2n}$ acting on the set $\pm 1,\dots,\pm n$, the transposition $(1,-1)$ results in a cut of the cycle
in which both $1$ and $-1$ are present; and $(1,-1)$ results in a union of the cycles
if $1$ is contained in one of these cycles and $-1$ is contained in the other (we saw this in the example above, in the case $n=1$). This action of transposition on the product of disjoint cycles in a permutation group is well known and easily verified.
 
 Next: the relation
 \be
 (1,-1)(2,-2)\cdots(n,-n)\otimes\frak{V}_1,\dots,\frak{V}_V=\frak{V}^*_1,\dots,\frak{V}^*_F
 \ee
 is nothing but the description of vertex monodromies of the graph $\Gamma$ and of the 
 vertex monodromies of the graph $\Gamma^*$. 
 
\subsection{Mixed ensembles $\Omega_{n_1,n_2,n_3}$}

The composed ensemble which consists of $n_1$ complex $N\times N$ matrices, of $n_2$ unitary
$N\times N$ matrices and
of $n_3$ matrices from $SU(N)$  with the measure 
which is the product of $n_1,n_2$ and $n_3$ measures on each group will be denoted by $\Omega_{n_1,n_2,n_3}$. We suppose that $n_1+n_2+n_3=n$ where $n$ is the number of ribbon edges 
of the embedded graph $\Gamma$.

Now we will consider multi-integrals over different groups, namely over $GL(N),U)N)$ and $SU(N)$:
$$
\langle f\rangle_{\Omega_{n_1,n_2,n_3}} =\int_{\Omega_{n_1,n_2,n_3}} f d\mu(X):=
\int_{GL(N)^{\times n_1}}\int_{U(N)^{\times n_2}}\int_{SU(N)^{\times n_3}}f(X)d\mu(X)
$$
where $X$ denotes the collection of $n=n_1+n_2+n_3$ random matrices which belong to $GL(N),U(N)$ or $SU(N)$ groups.
 
In this notations Proposition \ref{V=1} reads as
\be\label{V=1'}
\langle s_\lambda(\frak{V}) \rangle_{\Omega_{0,0,n}}=\left(d_{\lambda,U(N)}\right)^{-n}
\prod_{b=1}^F s_\lambda(\frak{V}^*_b)
\ee
and it is written for a graph $\Gamma$ with a single vertex.
 
\bp\label{mixed_equipment}
Suppose that each vertex monodromy dressed by random matrices contains at least one random matrix either from $GL(N)$ or from $U(N)$.
In this case for each $\Gamma$ and for any partition $\lambda$ we get
\be\label{sameSchurs}
\langle \prod_{a=1}^V s_{\lambda}\left(\frak{V}_a(X)\right)  \rangle_{\Omega_{n_1,n_2,n_3}}=\left(d_{\lambda,GN(N)}\right)^{-n_1} \left(\lambda,d_{U(N)}\right)^{-n_2-n_3}
\prod_{b=1}^F s_\lambda(\frak{V}^*_b),
\ee
\ep
Note that in contrast to (\ref{Schurs}) there is a single partition in the left hand side.

The proof follows from the fact that in (\ref{3bc}) in case $\lambda=\mu$ all terms vanish
in the left hand side except the term with $q=0$.

\section{Matrix models}

In addition to the set of monodromies we associate the semi-infinite set of
free parameters $\pb^{(a)}=\left(p^{(a)}_1,p^{(a)}_1,\dots \right)$, $a=1,\dots,V$
and 
$\tilde\pb^{(b)}=\left(\tilde p^{(b)}_1,\tilde p^{(b)}_1,\dots \right)$, $b=1,\dots,F$. 

For any matrix $Y$ and for every set $\pb=(p_1,p_2,\dots)$ we have Cauchy-Littlowood relation:
\be\label{CL}
e^{N\sum_{m=1}^\infty \frac 1m p_m \Tr(Y^m) }=
\sum_{\lambda} 
\prod_{a=1}^V s_\lambda(\pb)s_\lambda(Y)
\ee

Thanks to (\ref{CL}) and the proposition \ref{Propos}
in the case $G={U}(N)$, as well as in the case $G={GL}(N)$, we achieve
\bp
\be\label{prop}
Z_G(\pb,C):=
\int_{G^{\times n}} \prod_{a=1}^V e^{N\sum_{m=1}^\infty \frac 1m p^{(a)}_m \Tr\left(\frak{V}_a(X)  \right)^m }
d\mu(X)=
\sum_{\lambda} N^{-n|\lambda|}d_{\lambda,G}^{-n} 
\prod_{a=1}^V s_\lambda(N\pb^{a})
\prod_{b=1}^F s_\lambda(\frak{V}^*_b)
\ee
\be\label{prop^*}
\tilde Z_G(\tilde\pb,C):=
\int_{G^{\times n}} \prod_{b=1}^F e^{N\sum_{m=1}^\infty \frac 1m \tilde p^{(b)}_m  \Tr\left(\frak{V}^*_b(X)  \right)^m }
d\mu(X)=
\sum_{\lambda} N^{-n|\lambda|}d_{\lambda,G}^{-n} 
\prod_{b=1}^F s_\lambda(N\tilde\pb^{b})
\prod_{a=1}^V s_\lambda(\frak{V}_a)
\ee
\ep
(Here, $C$ denotes the collection of corner matrices).



As we see, the right-hand sides depend only on the spectrum of each of the monodromies.

\br
For a given $\Gamma$ we have a family of multimatrix models obtained by transformations of the set $\{ C_i,\, i=\pm 1,\dots,\pm n\}$ that preserve the set of vertex monodromies $\{\frak{V}^*_b,\,b=1,\dots,F\}$, or, more precisely, the set of spectra of these vertex monodromies (we will call such transformations gauge transformations).
\er

\subsection{Integrals over $G^{\times n}$, $G={SU}(N)$ \label{new}}

Now, we apply relations written down in Introduction to get solvable multi-matrix models
with $G={SU}(N)$.

Let's first consider the Brezin-Gross-Witten (BGW) matrix model \cite{GrossWitten}:
\be\label{GW}
Z_{U(N)}(A,B)=\int_{{U}(N)} e^{\beta\Tr(UA+U^\dag B)} d\mu(U)
\ee
which is the simplest example of the right hand side of (\ref{prop}) with $V=2,\,n=1,\,F=1$, where the graph $\Gamma$ is a segment drawn on $S^2$ with corner matrices $A$ and $B$.
In \cite{MirMorSemen} it was shown that it can be related to the KP tau function.
However, the integral
\be\label{GW-SU}
Z_{SU(N)}(A,B)=\int_{{SU}(N)} e^{\beta\Tr(UA+U^\dag B)} d\mu(U)
\ee
which was calculated in \cite{BChVol} contains an additional summation over $q$, which
enters into relation (\ref{3bc}) due to the Taylor expansion (\ref{CL}) of the two exponentials in the integrand, and
I cannot equate the integral (\ref{GW-SU}) to a tau function: it can be equated to a series of two-component KP tau functions, each labeled with a parameter $q$, see Appendix.

Under certain conditions for the graph $\Gamma$ the integration over $U(N)\times \cdots \times U(N)\times SU(N)\times\cdots\times SU(N)$
results in KP tau function.

However, in case we have a single exponential in the integrand we will use only relations 
(\ref{3'}) at the first step and both (\ref{4}) and (\ref{3'}) in all further steps leading to the 
${SU}(N)$ version of the relation \ref{prop}.

Which
means that only graphs with a single vertex is applicible: $V=1$.

\bp 
\be\label{new-prop}
Z(\pb,C):=
\int_{{SU}(N)^{\times n}} e^{N\sum_{m=1}^\infty \frac 1m p_m \Tr\left(\frak{V}(X)  \right)^m }
d\mu(X)=
\int_{{U}(N)^{\times n}} e^{N\sum_{m=1}^\infty \frac 1m p_m \Tr\left(\frak{V}(X)  \right)^m }
d\mu(X)=
\ee
\be
=\sum_{\lambda} N^{-n|\lambda|}d_{\lambda,G}^{-n} 
 s_\lambda(N\pb)
\prod_{b=1}^F s_\lambda(\frak{V}^*_b)
\ee
\ep

{Example 1} The only one-matrix model (it is the analogue of the Harish-Chandra-Itzykson-Zuber integral \cite{HarishChandra},\cite{ItzyksonZuber}):
\be\label{HCh-SU}
Z_{SU}:= \int_{SU(N)} e^{XAX^\dag B}d\mu(X)=\int_{U(N)} e^{XAX^\dag B}d\mu(X)=
\sum_{\lambda\atop\ell(\lambda)\le N} \frac{1}{(N)_\lambda}s_\lambda(A)s_\lambda(B)
\ee
where $d\mu(X)$ is the Haar measure in $SU(N)$ in the left integral and
is the Haar measure in $U(N)$ in the right integral and where 
\be
(N)_\lambda=\prod_{(i,j)\in\lambda}(N+j-i)
\ee
The right hand side of (\ref{HCh-SU}) is an example of hypergeometric tau function  \cite{OS-TMP}.

\paragraph{Composed multi-matrix model}

\bp
Under conditions of Proposition \ref{mixed_equipment}
we get
\be
\langle \prod_{a=1}^V e^{\sum_{m>0}\frac{N}{m}p_m^{(a)}\Tr(\frak{V}_a)^m}\rangle_{\Omega_{n_1,n_2,n_3}}=
\left(d_{\lambda,GL(N)} \right)^{-n_1}\left(d_{\lambda,U(N)} \right)^{-n_2-n_3}
\prod_{a=1}^V s_\lambda(N\pb^{(a)})
\prod_{b=1}^F s_\lambda(\frak{V}^*_b)
\ee
Here
\be
d_{\lambda,GL(N)}=s_\lambda(N\pb_\infty),\quad
d_{\lambda,U(N)}=s_\lambda(I_N)
\ee
\ep

\paragraph{Tau functions.}
To get tau functions we need to 
restrict ourselves to the case $V-n+F=2$ and to
impose some restrictions on our free parameters which
consist of the set $\pb$ and the set of dual monodromies $\{\frak{V}^*_b,\,b=1,\dots,n+1\}$.

The idea is to diminish the number of Schur functions in the enumarator
with the help of the known \cite{Mac} relations:

\be\label{restriction1}
 s_\lambda(\pb(u)) = s_\lambda(I_N)\prod_{(i,j)\in\lambda} \frac{u+j-i}{N+j-i}
,\quad \pb(u)=(u,u,u,\dots)
\ee
and
\be\label{restriction2}
s_\lambda(I[p]) = s_\lambda(I_N)\prod_{(i,j)\in\lambda} \frac{p+j-i}{N+j-i},
\ee
where the spectrum of a matrix $ I[p])$ consists of $N-p$ zeroes and $p$ units.
(Actually we have $s_\lambda(I[p])=s_\lambda(\pb(p))$.)

Thus, we need choose a number $u$ to equate some of $\pb$ to $\pb(u)$.
And to choose the numbers $p_b$ ($b=2,\dots,n+1$) and to equate the monodromies $\frak{V}^*_b$ to $I[p_b]$.
We want to get instead of the right hand side of (\ref{new-prop}) the series of type
\be\label{tau-hyp}
\sum_\lambda s_\lambda(\pb)s_\lambda(\frak{V}^*_1)\prod_{(i,j)\in\lambda} r(j-i)\quad 
{\rm or}\quad
\sum_\lambda s_\lambda(\frak{V}^*_2)s_\lambda(\frak{V}^*_2)\prod_{(i,j)\in\lambda} r(j-i)
\ee
where $r$ is a function on the lattice $Z$ which we call hypergeomtric tau function where
the product $\prod_{(i,j)\in\lambda} r(j-i)$ generalizes the Pochhammer symbol.
Sums of form (\ref{tau-hyp}) we called hypergeometric tau functions \cite{OS-TMP}.

To get it we need to choose $V-n+F=2$ and restrictions $n-1$ various restrictions of type 
(\ref{restriction1})-(\ref{restriction2}).

\paragraph{BKP tau function.}

Consider the following integral over $n$ integrals over $U(N)$ 
\be\label{SU-KP-BKP}
\int_{U(N)^{\times n}} 
\prod_{a=1}^{V-1} 
e^{N\sum_{m\ge 0}\frac{p_m}{m}\Tr\left(\frak{V}_{a} \right)^m}
\det\frac{1}{(I_N-\frak{V}_V\otimes \frak{V}_V)^{\frac12} }
\det\left(\frac{I_N+\frak{V}_V}{I_N-\frak{V}_V}\right)^{\frac 12} d\mu(X)=
\ee
\be
=\sum_{\lambda} d_{\lambda,G}^{-n-1} 
\prod_{a=1}^V s_\lambda(N\pb^{a})
\prod_{b=1}^F s_\lambda(\frak{V}^*_b)
\ee
This integral was evaluated in \cite{NO2020}.
Under restrictions (\ref{restriction1})-(\ref{restriction2}) this integral
is the BKP hypergeometric series which can be written as the pfaffian of a matrix whose
entries are generalized  hypergeometrc functions \cite{OST-I}.

Under certain conditions for the graph $\Gamma$ the integration over $U(N)\times \cdots \times U(N)\times SU(N)\times\cdots\times SU(N)$
resutlts in BKP tau function. It will be described in a more detailed text. In the present text 
we consider integrals over $G^{\times n}$ where $G$ is either $U(N)$ or is $SU(N)$.
This BKP tau function can be expressed as a Pfaffian \cite{OST-I}, see Appendix.

\section{Acknowledgements}

This work is an output of a research project implemented as part of the Basic Research Program at the National Research University Higher School of Economics (HSE University).

\bigskip  \bigskip

\appendix

\section{Hypergeometric tau functions}

$$
\partial_{p^{(1)}_1}\partial_{p^{(2)}_1} \phi_n=e^{\phi_{n-1}-\phi_n}-e^{\phi_{n}-\phi_{n+1}},
\quad \phi_n=\frac{\tau_n}{\tau_{n+1}}
$$
Let us fix $n$ and $\pb^{(2)}$. $\pb:=\pb^{(1)}$. Then
$$
3\partial_{p_3}\partial_{p_1} u = \frac 14 \partial_{p_1}^4 u + 3 \partial^2_{p_2}u
+\frac 34 (u^2)_{p_1},\quad u=2\partial^2_1\log\tau
$$

Hypergeometric tau functions solves not only bilinear but also linear equation,
which can be called string equations and which generalize Gauss equation for 
Gauss hypergeometric function ${_2}F_1$.

There are few determinantal representations of these tau functions.
For the various properties of these tau functions see \cite{OS2000},\cite{OS-TMP}.

\section{Fermionic construction of 2KP tau functions related to (\ref{GW-SU})}

Using (\ref{CL}) and (\ref{3bc}) we get
\be
Z_{SU(N)}(A,B)=\sum_{q\in Z}\tau(q,\pb^{(1)},\pb^{(2)})
\ee
where
\be
\tau(q,\pb^{(1)},\pb^{(2)})=\langle N,-N|\prod_{a=1,2}e^{\sum_{m>0} \frac 1m p^{(a)}J_1^{(a)}} e^{\sum_{i\in Z} a_i\psi^{(1)}_i\psi^{\dag(2)}_{i+q}}
|0,0\rangle
\ee
where $\pb^{(a)}=(1,0,0,\dots)$ and
$$
a_i=\begin{cases}
     i>0\\
     i<0
    \end{cases}
$$

\section{Derivation of (\ref{2'}) for complex $A$ and $B$. \label{derivation}}

In a number of papers (for instance in \cite{AOV},\cite{2DYM}) the following relation was used:

\bp
 Let $A$ and $B$ be $N\times N$ complex matrices. Then
\be\label{Prop1}
C_N\int_{\mathbb{U}_N} s_\lambda(UAU^\dag B) d_*U=\frac{s_\lambda(A)s_\lambda(B)}{s_\lambda(I_N)}
\ee
where $d_*U$ is he Haar measure on $\mathbb{U}_N$ and $C_N$ is chosen from the normalization
$C_N\int_{\mathbb{U}_N} d_*U=1$ (and in what follows we include it to the definition of $d_*U$) and where $s_\lambda$ is the Schur function labeled by
a partition $\lambda$, see \cite{Mac}.
\ep

I did not see the written proof of (\ref{Prop1}) and I present it below.
The proof consists of few simple steps (i)-(iii):

\,

(i) Firstly, we note that for $A,B\in\Sigma $, where $\Sigma$ is the space of Hermitian matrices
the identity (\ref{Prop1}) is known, see
 Example 3  in Section 5 of Chapter VII in the textbook \cite{Mac}, the page 445 (and also page 441 where $\Sigma$ is defined).

 (ii) Then, due to the invariant property of the Haar measure we can replace $A$ and
 $B$ by diagonal matrices of eigenvalues $X=\diag(x_1,\dots,x_N)$ and $Y=\diag(y_1,\dots,y_N)$ respectively, where $x_a,y_a$ are real.  Then, since the both sides of (\ref{Prop1}) are polynomial in each variable
 $x_a,y_a$ ($a=1,\dots,N$) we can analiticaly continue both sides for complex $x_a,y_a$. Thus,
 we get
 $$
\int_{\mathbb{U}_N} s_\lambda(UXU^\dag Y) d_*U=\frac{s_\lambda(X)s_\lambda(Y)}{s_\lambda(I_N)}
 $$
 where $X$ and $Y$ are complex diagonal\footnote{Using the Cauchy-Littlewood relation the this was used in \cite{O-new2003} to re-derive
 the determinantal answer for the HCIZ integral \cite{HarishChandra},\cite{ItzyksonZuber}
 $\int e^{\tr UAU^\dag B}d_*U$
 for normal matrices $A$ and $B$.}.
 
 Consider strictly upper-triangular matrices
\be\label{Delta}
\Delta^{(A)}_{a,b}=\Delta^{(B)}_{a,b}=0,\quad a\ge b.
\ee

Suppose that $\tilde A$ and $\tilde B$ have the form
\be\label{Schur-decomposition}
\tilde A=U_1(X+\Delta^{(A)})U_1^\dag,\quad
\tilde B=U_2(Y+\Delta^{(B)})U_2^\dag,\qquad
U_i\in \mathbb{U}_N,\quad i=1,2
\ee
where $X=\diag\{x_1,\dots,x_N\}$ and $Y=\diag\{y_1,\dots,y_N\}$ are complex diagonal matrices, and $\Delta^{(i)}$ are strictly upper-triangular, see (\ref{Delta}).
Now $\tilde A$ and $\tilde B$ are arbitrary complex matrices and (\ref{Schur-decomposition})
are their Schur decomposition.
We have $s_\lambda(\tilde A)=s_\lambda(X)$ and $s_\lambda(\tilde B)=s_\lambda(Y)$.
Let us prove that
\be
\int_{\mathbb{U}_N} s_\lambda(UAU^\dag B) d_*U=
\int_{\mathbb{U}_N} s_\lambda(UXU^\dag Y) d_*U.
\ee

To verify it let us use a result of B.Collins (see \cite{Collins} and \cite{Collins2}).
Let
\be
{\cal U}_N({\bf a},{\bf a}',{\bf b},{\bf b}'):=
 U_{a_1,b_1}\cdots U_{a_c,b_c}(U^\dag)_{b'_1,a'_1}\cdots (U^\dag)_{b'_{c'},a'_{c'}}
\ee
which is a monomial in the entries of matrices $U\in\mathbb{U}_N$ and it's inverse one
labeled, respectively, by sets ${\bf a},{\bf b}$ and ${\bf a}',{\bf b}'$.

Then, the nice result of \cite{Collins} is
\be\label{collins}
\int_{\mathbb{U}_N}{\cal U}_N({\bf a},{\bf a}',{\bf b},{\bf b}')d_*U
=\delta_{c,c'}\sum_{\sigma,\tau\in S_c}{Wg}_N(\tau\sigma^{-1})\prod_{k=1}^c\delta_{a_k,a'_{\sigma(k)}}\delta_{b_k,b'_{\tau(k)}}
\ee
where $S_c$ is the symmetric group and $Wg_N$ is some function on $S_c$. Let us note that the integral in the left hand side
of (\ref{collins}) was also evaluated in other works, see \cite{Morozov-U-int},\cite{Shatashvili}, however it is (\ref{collins}) we need in.
$Wg_N$ is called Weingardten's function,
we do not need it's explicit form.
In view of (\ref{collins}) in case $c=c'$ it is natural to introduce two balance numbers
(or, balance functions defined on monomials terms where $c=c'$)
\be
{\cal A}\left({\cal U}_N({\bf a},{\bf a}',{\bf b},{\bf b}')\right)=\sum_{i=1}^c a_i-\sum_{i=1}^{c} a'_i
\ee
and
\be
{\cal B}\left({\cal U}_N({\bf a},{\bf a}',{\bf b},{\bf b}'\right)=\sum_{i=1}^{c} b_i-\sum_{i=1}^{c} b'_i.
\ee

In what following we omit the dependence of $N,{\bf a},{\bf a}',{\bf b},{\bf b}'$ and write
just ${\cal U}$.
From (\ref{collins}) we have obvious

\bl\label{vanishing}
 ${\cal U}$ necessarily vanishes in one of the following cases

(A) ${\cal A}({\cal U})\neq 0$ ($c=c'$)

(B) ${\cal B}({\cal U})\neq 0$  ($c=c'$)

(C) $c\neq c'$.

\el

The last condition (C) results in the presence of $\delta_{q,0}$ in the right hand side 
of (\ref{2'}).

(iii) Next, we remind that the Schur function admit the following Taylor series
\be\label{char-map}
s_\lambda(X)=\sum_{\Delta} \phi_\lambda(\Delta)\tr\left( X \right)^{\Delta_1} \cdots
\tr\left( X \right)^{\Delta_{\ell(\Delta)}}
\ee
which is called character map relations \cite{Mac} because $s_\lambda$ is the character of
the linear group and $\phi_\lambda$ are specially normalized characters of the symmetric group.
The right hand side consists of the sum of monomials of the entries of $X$.

Suppose $X=UAU^\dag B$ and the matrices $A$ and $B$ are both diagonal. Then, as one can notice
the integral over $U_N$ of each term which is monomial in the entries $X$ (this follows from the specific form of such monomials arising from traces and from the diagonality of $A$ and $B$.
For instance, $\tr UAU^\dag B = \sum U_{a,b}A_{b,b'}(U^\dag)_{b',a'}B_{a',a}$ with $a=a',\,b=b'$.)

Now suppose that $A$ is a diagonal plus a given strictly upper triangle matrix $\Delta^{(A)}$:
$\Delta^{(A)}_{i,j}=0,\,i\ge j$. For $j-i>0$ we call this number a content of the entry $\Delta^{(A)}_{i,j}$.   Then any polynomial originated from a factor
$\tr\left(UAU^\dag B\right)^m$ which contain a factor $\Delta^{(A)}_{i,j},\,i<j$ diminish
the balance number ${\cal B}$ by $j-i>0$. Each entry of $\Delta^{(A)}$ which enters a monomial ${\cal U}$ diminishes the balance number ${\cal B}$ by its content. Therefore, according to the (B) there is no contribution of $\Delta^{(A)}$
to the integral of each $\tr\left(UAU^\dag B\right)^{\mu_1}\cdots \tr\left(UAU^\dag B\right)^{\mu_k}$ and therefore, no contribution to the integral (\ref{Prop1}). Notice, that
the matrix $\Delta$ does not change the balance number ${\cal A}$.

Simirlarly, suppose the matrix $B$ is a diagonal matrix plus a strictly upper triangle matrix $\Delta^{(B)}$: $\tilde\Delta^{(B)}_{i,j}=0,\,i\ge j$. In this case, each monomial, say, ${\cal U}$  which includes any number of entries of $\Delta^{(B)}$ has the balance number ${\cal A}$ enlarges by positive number (equal to the sum of contents of the entries) and according to (A) does not contribute to the integral.

We also note that terms with $\Delta^{(A)}$ does change the balance number ${\cal A}$ and
terms with $\Delta^{(B)}$ does not change the balance number ${\cal B}$.

Since, any complex matrix can be presented in form of diagonal plus strictly triangle by the
conjugation by unitary ones (the Schur decomposition) and also due to the invariance property of the Haar measure we
obtain the proof of (\ref{Prop1}) for any complex $A$ and $B$.

\end{document}